\documentclass[a4paper]{jpconf}
\usepackage{graphicx}
\usepackage{amsmath}
\usepackage{amssymb}
\usepackage{color}

\begin{document}
\title{The generalized Kadanoff-Baym ansatz. Computing nonlinear response properties of finite systems}

\author{K.~Balzer}
\address{University of Hamburg, Max Planck Research Department for Structural Dynamics, Building 99 (CFEL), Luruper Chaussee 149, 22761 Hamburg, Germany}
\ead{karsten.balzer@mpsd.cfel.de}

\author{S.~Hermanns and M.~Bonitz}
\address{University of Kiel, Institut f\"ur Theoretische Physik und Astrophysik, Leibnizstrasse 15, 24098 Kiel, Germany}

\newcommand{\eq}[1]{Eq.~(\ref{#1})}
\newcommand{\eqsand}[2]{Eqs.~(\ref{#1}) and~(\ref{#2})}
\newcommand{\eqsthree}[3]{Eqs.~(\ref{#1}), (\ref{#2}) and~(\ref{#3})}
\newcommand{\ii}{\mathrm{i}}
\newcommand{\mean}[1]{\langle #1 \rangle}
\newcommand{\cc}{{\cal C}}
\newcommand{\dc}{\delta_{\cal C}}
\newcommand{\tc}{T_{\cal C}}
\newcommand{\op}[1]{\hat{#1}}
\newcommand{\nn}{\nonumber}
\newcommand{\mret}{{\mathrm{ret}}}
\newcommand{\madv}{{\mathrm{adv}}}
\newcommand{\bra}[1]{\langle{#1}|}
\newcommand{\ket}[1]{|{#1}\rangle}
\newcommand{\braket}[2]{\langle{#1}|{#2}\rangle}
\newcommand{\thc}{\theta_{\cal C}}
\newcommand{\intc}[1]{\int_{\cal C}\mathrm{d}{#1}\;}
\newcommand{\intlim}[3]{\int_{#1}^{#2}\mathrm{d}{#3}\;}
\newcommand{\todo}[1]{\textcolor{red}{\underline{#1}}}

\renewcommand{\e}[1]{\mathrm{e}^{#1}}

\begin{abstract}
For a minimal Hubbard-type system at different interaction strengths $U$, we investigate the density-response for an excitation beyond the linear regime using the generalized Kadanoff-Baym ansatz (GKBA) and the second Born (2B) approximation. We find strong correlation features in the response spectra and establish the connection to an involved double excitation process. By comparing approximate and exact Green's function results, we also observe an anomalous $U$-dependence of the energy of this double excitation in 2B+GKBA. This is in accordance with earlier findings [K.~Balzer \emph{et al.}, EPL~\textbf{98}, 67002 (2012)] on double excitations in quantum wells.
\end{abstract}

\section{Introduction}
The study of strongly correlated quantum systems and materials, e.g.,~\cite{pavarini11} and references therein, is a very rapidly evolving field in both experimental and theoretical physics. Especially the out-of-equilibrium dynamics is of great current interest in solid-state, atomic and molecular physics, in nanoelectronics, quantum transport etc. In all these fields, the availability of intense and coherent radiation, combined with ultra-short laser pulses, has triggered many key experiments that allow one to investigate matter under extreme nonequilibrium conditions where correlation and nonlinear effects are important. As examples, consider the photoionization of multi-electron atoms and molecules~\cite{becker12}, the many-body dynamics of particles in optical lattices~\cite{bloch08} or quantum interference effects in Mott insulators~\cite{wall11}.

In the recent decade, the nonequilibrium Green's function (NEGF) approach has developed into a powerful numerical tool and is able to address the dynamics of very different quantum systems. Examples include the excitation of non-ideal (semiconductor) plasmas~\cite{kwong98,kwong00}, nuclear collisions~\cite{rios11}, the carrier dynamics and carrier-phonon interaction in quantum dots, wells and lattice networks contacted to leads~\cite{gartner06,lorke06,uimonen11,khosravi12}, and the problem of baryogenesis in cosmology~\cite{garny11}. In all these cases, the central quantity is the one-particle nonequilibrium Green's function defined on the complex Keldysh contour $\cc$,
\begin{align}
\label{gdef}
 g_{ss'}^\sigma(t,t')&=-\ii\mean{\tc\,\op{c}_{s,\sigma}(t)\,\op{c}^\dagger_{s',\sigma}(t')}\\
 &=\thc(t-t')\,g_{ss'}^{\sigma,>}(t,t')+\thc(t'-t)\,g_{ss'}^{\sigma,<}(t,t')\;,\nn
\end{align}
with non-temporal degrees of freedom $s$, $s'$ and $\sigma$. For our purpose, $s$ and $s'$ will refer to spatial degrees of freedom, and $\sigma=\uparrow,\downarrow$ will indicate the spin, assuming diagonality in the spin subspace. Further, the operator $\tc$ accounts for contour ordering of the times $t$ and $t'$, and $\mean{ \ldots}$ means averaging in the grand canonical ensemble.

In principle, the two-time dependence of the Green's function enables the systematic and exact treatment of dynamic correlation effects independently of the strength of the interaction and regardless of the presence of weak or strong external fields. Even interaction quenches can be studied, e.g.~\cite{eckstein10}. However, the application to large system sizes and the investigation of the system's long time behavior usually demand additional approximations and simplifications to the equations of motions of the NEGF~\cite{book_bonitz_qkt}. 

The most straightforward approach is to perform a many-body approximation (MBA) on the self-energy. Due to diagram expansions, this can be done in a systematic and conserving way and leads, e.g., to the Hartree-Fock (HF), second(-order) Born (2B), GW or T-matrix approximations. Correlation effects are then partially treated beyond the level of Hartree-Fock. Additional simplifications typically concern the two-time structure which renders NEGF calculations demanding~\cite{balzer10_pra1,balzer10_pra2}. Within the generalized Kadanoff Baym ansatz (GKBA)~\cite{lipavsky86,hermanns12_pysscripta}, it is possible to study the equations of motion for the Green's function in the single-time limit, which drastically reduces the numerical effort.

In general, the application of MBAs and other simplifications such as the GKBA require thorough justification as they can lead to unphysical effects such as self-interaction errors and spurious dynamical excitations~\cite{hermanns12_prb}, bistability~\cite{khosravi12} or artificial steady states~\cite{puigvonfriesen10}. For this reason, we, in this contribution, extend previous work on weak perturbations~\cite{balzer12_epl}, which included the discussion of double excitations (see also~\cite{sakkinen12}), to the nonlinear regime and analyze the performance of the GKBA for a finite system with the two-body interactions being treated in the second Born approximation.

\section{Equations of motion and GKBA}
Introducing the one-particle self-energy $\Sigma^\sigma_{ss'}(t,t')$, the Green's function (\ref{gdef}) obeys the Kadanoff-Baym equation (KBE)~\cite{book_kadanoffbaym_qsm},
\begin{align}
\label{kbe}
 \left(-\ii\frac{\partial}{\partial t}\delta_{s\bar{s}}-h^\sigma_{s\bar{s}}(t)\right)g_{\bar{s}s'}^\sigma(t,t')=\dc(t-t')\delta_{ss'}+\intc{\bar{t}}\Sigma^\sigma_{s\bar{s}}(t,\bar{t})g_{\bar{s}s'}^\sigma(\bar{t},t')\;,
\end{align}
and its adjoint with $t\leftrightarrow t'$. Here, $h^\sigma_{ss'}$ denotes the single-particle (kinetic plus potential) energy, and summation over $\bar{s}$ is implied.

Using the generalized Kadanoff-Baym ansatz, we transform the set of KBEs into a single equation for the reduced density matrix $\rho_{ss'}^{\sigma,<}(t)=-\ii g_{ss'}^{\sigma,<}(t,t)$ or equivalently $\rho_{ss'}^{\sigma,>}(t)=-\ii g_{ss'}^{\sigma,>}(t,t)$ retaining time causality and important conservation laws such as the conservation of total energy, momentum and density. In the course of this, the time-off-diagonal components of the NEGF are reconstructed as (we again sum over $\bar{s}$), 
\begin{align}
\label{gkbadef}
 g_{ss'}^{\sigma,\gtrless}(t,t')=-g^{\sigma,\mret}_{s\bar{s}}(t,t')\,\rho_{\bar{s}s'}^{\sigma,\gtrless}(t')+\rho_{s\bar{s}}^{\sigma,\gtrless}(t)\,g^{\sigma,\madv}_{\bar{s}s'}(t,t')\;,
\end{align}
and inserted into the time-diagonal collision integral (the r.h.s.~of the KBEs for $t=t'$ in the second Born approximation). For an explicit expression of $\Sigma^\sigma_{ss'}(t,t')$ as functional of the NEGF and for the corresponding collision term, see, e.g., the contribution of S.~Hermanns \emph{et al.} in the present conference proceedings. Finally, the treatment of the two-time retarded and advanced propagators $g^{\sigma,\mret}_{ss'}(t,t')$ and $g^{\sigma,\madv}_{ss'}(t,t')$ in Hartree-Fock approximation renders the ansatz practical:
\begin{align}
 g^{\sigma,\mret/\madv}_{ss'}(t,t')=\left.\mp\ii\thc(\pm[t-t'])\exp{\left(-\ii\intlim{t'}{t}{\bar{t}}h^\sigma_\mathrm{HF}(\bar{t})\right)}\right|_{ss'}\;.
\end{align}
Here, $h^\sigma_\mathrm{HF}(t)$ denotes the single-particle time-dependent Hartree-Fock Hamiltonian.

We emphasize that the reconstruction of the greater and lesser components of the Green's function with real $t$ and $t'$ is sufficient as long as the method of adiabatic switching is applied to generate the correlated initial (ground) state by time propagation, for details see \cite{hermanns12_pysscripta}.

\section{Nonlinear density-response}
As outlined above, we are interested in the dynamical behavior of a finite quantum system beyond the regime of linear response. To this end, we consider a one-dimensional Hubbard model at half-filling with hopping $T$ and on-site interaction $U$ and strongly perturb it by changing the site energies in time~\cite{puigvonfriesen09}.

The initial Hamiltonian for times $t<0$ reads,
\begin{align}
\label{h0}
 \op{H}_{0}=-T\sum_{<s,s'>}\sum_{\sigma=\uparrow,\downarrow}\op{c}^\dagger_{s,\sigma}\op{c}_{s',\sigma}+U\sum_{s}\op{n}_{s,\uparrow}\,\op{n}_{s,\downarrow}\;,
\end{align}
where $s$ and $s'$ range from $0$ to $L-1$ for a chain of length $L$, and $<\!s,s'\!>$ indicates nearest-neighbor sites. Further, $\op{n}_{s,\sigma}=\op{c}^\dagger_{s,\sigma}\op{c}_{s,\sigma}$ denotes the density operator, and the energy (time) is measured in units of $T$ (the inverse hopping $T^{-1}$). Generally, we study the chain for open boundary conditions but this is irrelevant for the special case of two sites.

The system is excited by an instantaneous change of the site energies from zero to a positive finite value $\epsilon_s$. If two or more sites differ in energy after the switching process, a correlated electron motion is initiated in the chain. For simplicity, we detune only the energy of the first site which leads to the perturbation,
\begin{align}
\label{h1}
 \op{H}_1=\epsilon_0\,\theta(t)\sum_{\sigma=\uparrow,\downarrow}\op{n}_{0,\sigma}\;.
\end{align}
For $t>0$ and arbitrary $\epsilon_0$, the perturbed system $\op{H}_0+\op{H}_1$ will initially show a depopulation of the first site followed by an accumulation of density on the second. Subsequently, also the density on the remaining sites will change with time, and, finally, all occupations will start to oscillate. In the case of $\epsilon_0\ll1$, the population change will be small, such that the dynamics should be well characterized by the linear response properties of the chain. For $\epsilon\gtrsim1$, however, we expect nonlinear effects to be of importance. In the following, we consider the case $\epsilon_0=5.0$ and resort to the zero-temperature limit $\beta\rightarrow\infty$. Initially, at $t=0$, the system is prepared in the ground state of $\op{H}_0$.

We measure the response of the Hubbard chain with respect to the time-dependent electron density on the first site ($s=0$),
\begin{align}
\label{gammadef}
 \gamma^\sigma(t)=\mean{\op{n}_0^\sigma}(t)-\frac{1}{t^*}\intlim{0}{t^*}{\bar{t}}\mean{\op{n}_0^\sigma(\bar{t})}\;,\\
 \label{gammadefft}
  \gamma^\sigma(\omega)=\intlim{0}{t^*}{t}\exp(-\ii\omega t)\,\gamma^\sigma(t)\;,
\end{align}
where $\mean{\op{n}_0^\sigma}(t)=-\ii g_{00}^{\sigma,<}(t,t)$. Due to the spin symmetry obeyed by \eq{h1}, it is $\gamma(\omega)=\gamma^\uparrow(\omega)=\gamma^\downarrow(\omega)$. Furthermore, the time $t^*$ indicates a finite propagation time used in the numerics and is chosen sufficiently long such that it only affects the basic width of the peaks in $\gamma(\omega)$ but not their position.

\subsection{Exact dynamics}
In order to benchmark the approximate solution of the KBEs under the use of the GKBA and the second Born approximation, we compare to exact reference data computable for small $L$. The exact electron dynamics are obtained from propagating the one-particle nonequilibrium Green's function independently of the KBEs. To this end, we rewrite the average in the definition (\ref{gdef}) of the NEGF as,
\begin{align}
\label{gdefexact}
 g_{ss'}^\sigma(t,t')=-\frac{\ii}{Z_0}\sum_{N_\uparrow,N_\downarrow=0}^{L}\sum_{n}\e{-\beta(E_{N_\uparrow,N_\downarrow;n}-\mu N)}\bra{N_\uparrow,N_\downarrow;n}\,\tc\,\op{c}_{s,\sigma}(t)\,\op{c}^\dagger_{s',\sigma}(t')\,\ket{N_\uparrow,N_\downarrow;n}\;,
\end{align}
where $Z_0=\sum_{N_\uparrow,N_\downarrow}\sum_n\e{-\beta(E_{N_\uparrow,N_\downarrow;n}-\mu N)}$ is the partition function, $N=N_\uparrow+N_\downarrow$, and $n$ labels the states in the $N$-particle subspace. For small system sizes, the action of the time-dependent creation and annihilation operators,
\begin{align}
\label{ucudef}
\op{c}_{s,\sigma}^{(\dagger)}(t)&=\op{U}(0,t)\,\op{c}_{s,\sigma}^{(\dagger)}\,\op{U}(t,0)\;,
\end{align}
with,
\begin{align}
\label{udef}
\op{U}(t',t)&=\exp\left(-\ii\intlim{t}{t'}{\bar{t}}(\op{H}(\bar{t})-\mu \op{N})\right)\;,
\end{align}
on the energy eigenstates $\ket{N_\uparrow,N_\downarrow;n}$ (with energy $E_{N_\uparrow,N_\downarrow;n}$) can be computed numerically exactly. For each particle sector involved in \eq{gdefexact}, we work in the trivial site basis which involves the states $\ket{0}$, $\ket{\uparrow}$, $\ket{\downarrow}$ and $\ket{\uparrow\downarrow}$ and apply the Krylov method~\cite{hochbruck97} together with a high-order commutator-free exponential time propagation, see~\cite{alvermann11}.

\begin{figure}[t]
 \includegraphics[width=0.9\textwidth]{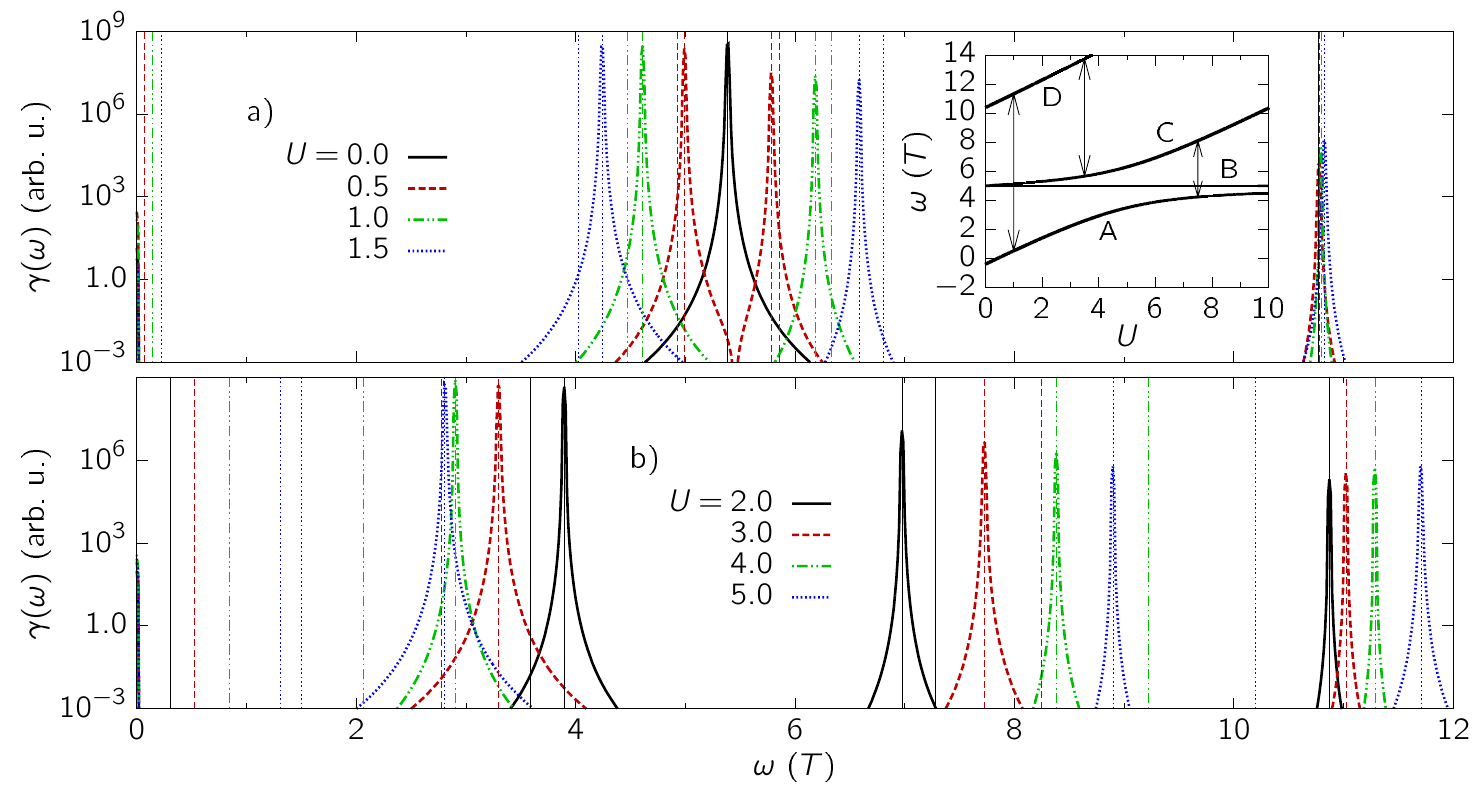}
 \caption{Exact nonlinear density-response spectra $\gamma(\omega)=\gamma^\uparrow(\omega)=\gamma^\downarrow(\omega)$ for a two-site chain at $\epsilon_0=5.0$ and half-filling. a) The noninteracting case $U=0$ and low to moderate interaction strengths $U=0.5$, $1.0$ and $1.5$. b) Moderate towards strong interaction $U=2.0$, $3.0$, $4.0$ and $5.0$. Inset in a): Static energy spectrum of the asymmetric two-site chain with $\epsilon_0=5.0$ and $\epsilon_1=0$. From left to right, the arrows indicate the transition frequencies $\omega_\mathrm{AD}$, $\omega_\mathrm{CD}$ and $\omega_\mathrm{AC}$.}
 \label{fig1}
\end{figure}

In Figure~\ref{fig1}~a) and b), we show exact results based on \eqsthree{gdefexact}{ucudef}{udef}, where the consideration of the thermodynamic ground state limits the trace in \eq{gdefexact} to $N_\uparrow=N_\downarrow=1$. The thick curves show the response $\gamma(\omega)$ for a two-site chain ($L=2$) at different repulsive interaction strengths $U$ according to \eq{gammadefft}.  For $U=0$, see the black curve in the upper panel, there exist two peaks---one at $\omega_0=5.385$ and one at $2\omega_0=10.770$. Interestingly, for $U>0$,  the energetically lowest peak at $\omega_0$ develops into a double-peak structure of which the left part steadily has a larger spectral weight. For larger $U$ there is even a weight difference of several orders of magnitude, cf.~Figure~\ref{fig1}~b). On the contrary, the peak at $2\omega_0$ changes only slightly with $U$ and almost maintains its spectral weight. 

To understand the spectrum, we recall the switch between two time-independent Hamiltonians given by \eqsand{h0}{h1} without a ramp function. In this context, the initial state (given by the ground state of $\op{H}_0$) can be expressed as a superposition of eigenstates of $\op{H}_{2}=\op{H}_0+\op{H}_1$ for $t>0$. Consequently, the transition frequencies between the eigenstates of $\op{H}_{2}$ determine the system's nonequilibrium dynamics, and we expect them therefore to appear in $\gamma(\omega)$. The inset in Figure~\ref{fig1}~a) shows the eigenenergies in the asymmetric chain with $\epsilon_0=5.0$ and $\epsilon_1=0$ as a function of $U$, and the thin vertical lines in Figure~\ref{fig1}~a) and b) indicate the associated excitation frequencies for the selected interaction strengths. Some of the vertical lines indeed coincide with the peaks in $\gamma(\omega)$. More precisely, we find three out of six transitions: $\omega_\mathrm{AC}$, $\omega_\mathrm{CD}$ and $\omega_\mathrm{AD}$, see the 
arrows in the inset. In the limit $U\rightarrow0$, $\omega_\mathrm{AC}$ and $\omega_\mathrm{CD}$ become degenerate which explains the double-peak structure for small, finite $U$.

\begin{figure}[t]
 \includegraphics[width=0.9\textwidth]{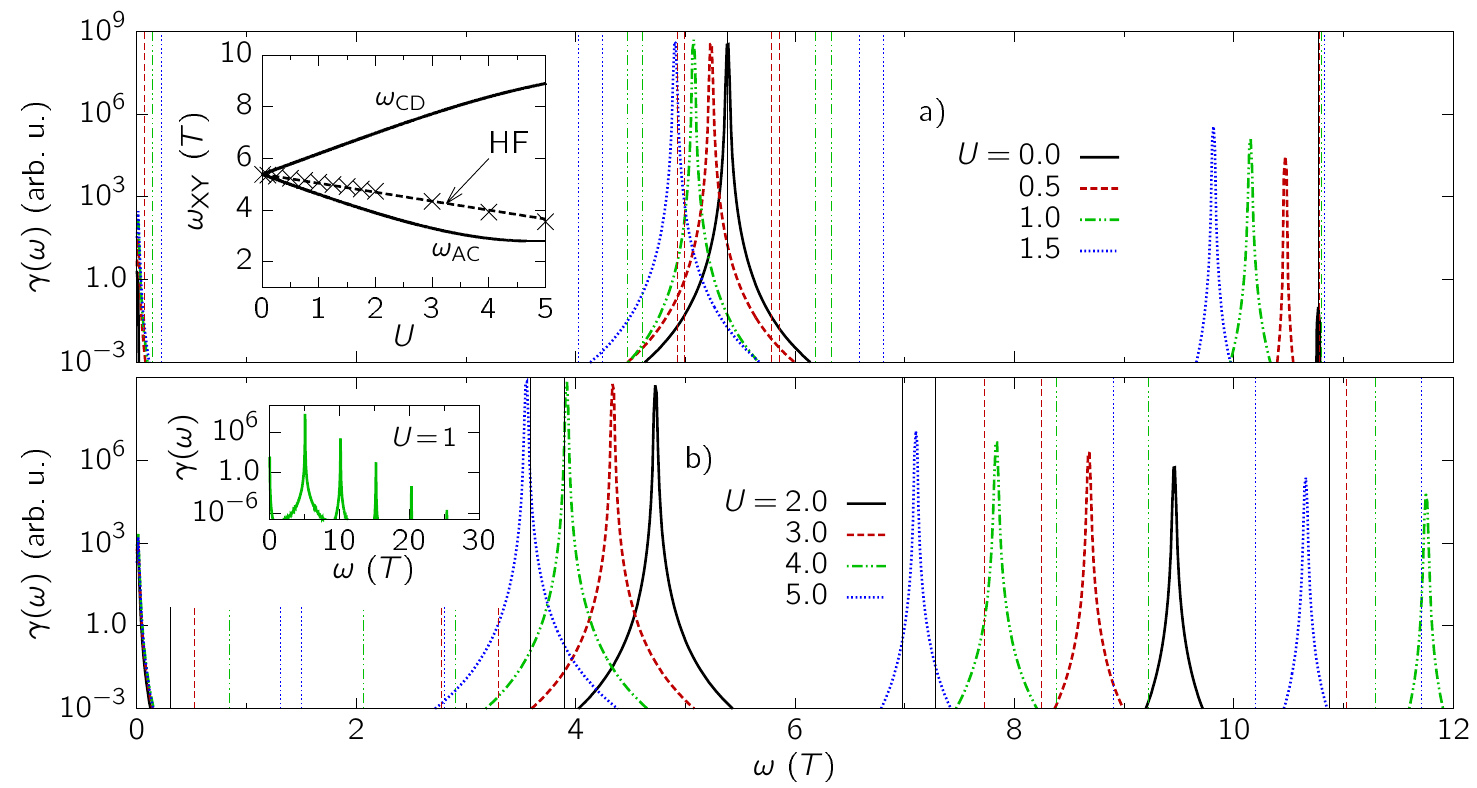}
 \caption{a) and b): Nonlinear density-response spectra $\gamma(\omega)$ in Hartree-Fock (HF) approximation. The parameters and the thin vertical lines indicating the exact transition frequencies are the same as in Figure~\ref{fig1}. The inset in a) compares the $U$-dependence of the exact transition frequencies $\omega_{\mathrm{AC}}$ and $\omega_{\mathrm{CD}}$ (solid lines) to the Hartree-Fock result (crosses). The dashed line is a linear fit, cf.~also Figure~\ref{fig5} and \eq{fithf}. For $U=1.0$, the inset in b) shows the emergence of ``higher harmonics'' of the main peak in the HF solution.}
 \label{fig2}
\end{figure}

\subsection{Approximate dynamics}
In this section, we investigate the density-response $\gamma(\omega)$ in Hartree-Fock as well as in second Born approximation propagating the KBEs~(\ref{kbe}). In the latter case, we in addition apply the GKBA of \eq{gkbadef}.

The results of Hartree-Fock (HF) calculations for the two-site chain are summarized in Figure~\ref{fig2} a) and b). What immediately becomes evident is the fact that there is no double-peak structure developing for finite $U>0$ as in Figure~\ref{fig1}. Instead, we observe only a single peak below $\omega=6$ which moves towards smaller energies as $U$ is increased. The inset in the upper panel indicates that the corresponding peak position changes linearly with the on-site interaction $U$. Moreover, its proximity to the lower black solid line referring to $\omega_{\mathrm{AC}}$ as a function of $U$ suggests that the HF approximation covers only the transition $\mathrm{A}\leftrightarrow\mathrm{C}$, compare with Figure~\ref{fig1}~a) (inset). On top of that, the HF approximation seems not  to describe the structure of $\gamma(\omega)$ for frequencies around $\omega=11$. Here, the observed peaks are obviously ``higher harmonics'' of the main peak, cf.~the inset in Figure ~\ref{fig2}~b).

\begin{figure}[b]
 \includegraphics[width=0.9\textwidth]{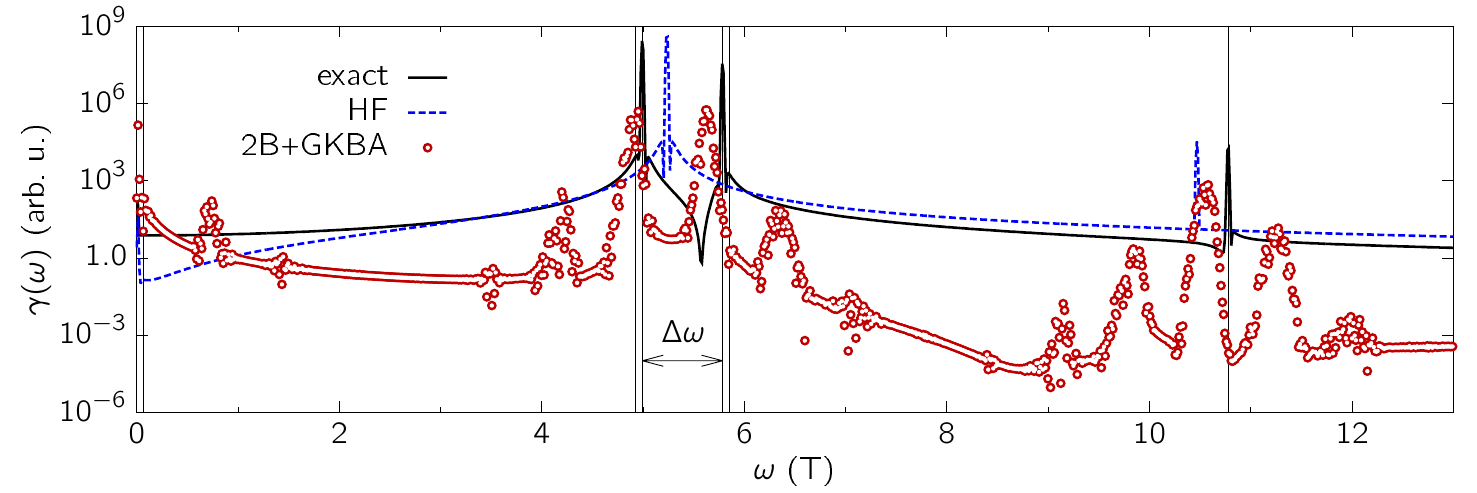}
 \caption{Dynamics of the two-site chain as in Figures~\ref{fig1} and \ref{fig2} for the case of $U=0.5$. Comparison of the exact density response $\gamma(\omega)$ to the result in Hartree-Fock (HF) and second Born (2B+GKBA) approximation. The thin black vertical lines are the same as the red dashed lines in Figures~\ref{fig1}~a) or~\ref{fig2}~a).
 }
 \label{fig3}
\end{figure}

Next, we perform GKBA calculations including correlations on the second Born (2B) level. Figure~\ref{fig3} shows the result (dots) for $U=0.5$ and compares to the HF solution (dashed line) and the exact data (solid line). Interestingly, the 2B+GKBA calculation drastically improves the HF result in the sense that, instead of a single peak at $\omega_\mathrm{HF}=5.234$, there appears a double-peak structure with practically vanishing spectral weight around $\omega_\mathrm{HF}$ and almost the correct energy difference. However, the individual peaks are broadened~\cite{broadening}, and in addition there emerge side peaks at distances comparable to the double-peak energy spacing $\Delta\omega$. Furthermore, we obtain a similar structure for $\gamma(\omega)$ between $\omega=9$ and $12$ but with one central peak. Importantly, the position of this peak is not a multiple of the frequency of one of the peaks belonging to the double-peak at lower frequencies. Hence, it is not a higher harmonics artifact as observed in 
the Hartree-Fock solution (see also Figure~\ref{fig5}).

\begin{figure}[t]
 \includegraphics[width=0.9\textwidth]{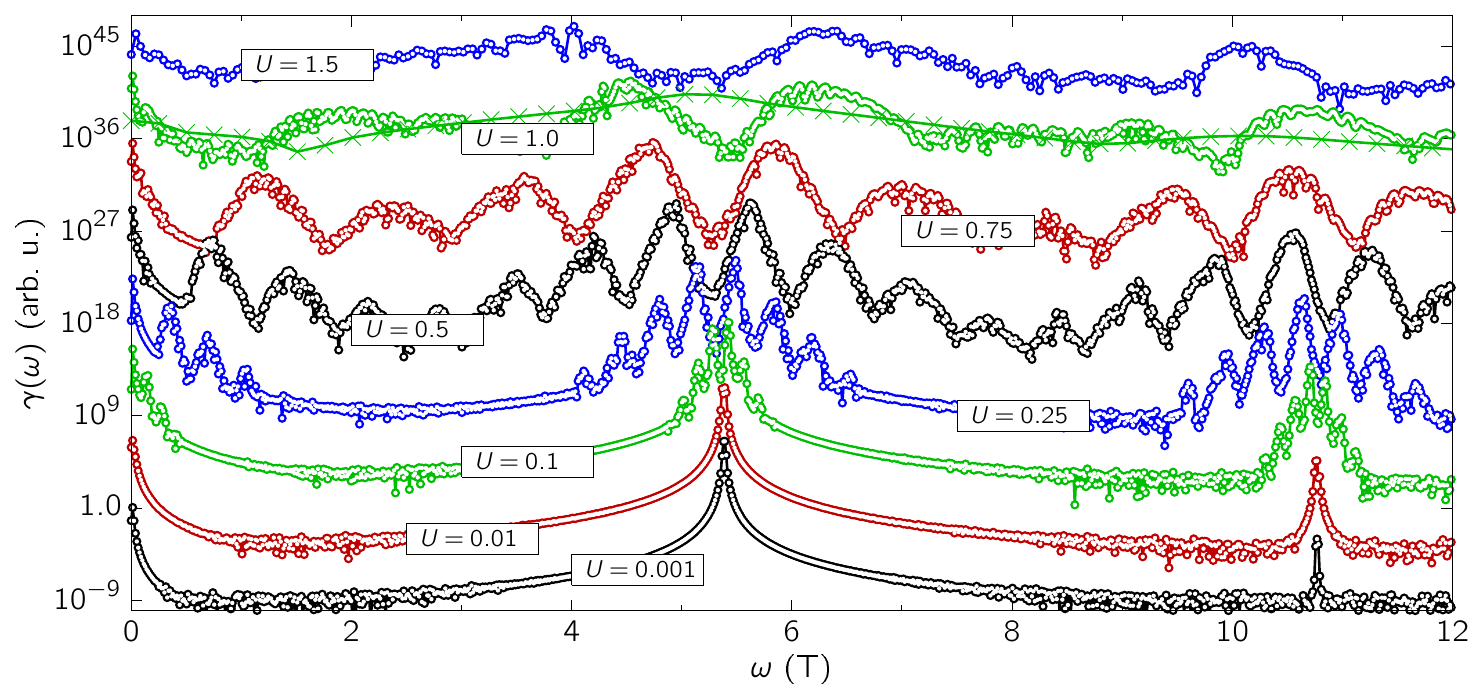}
 \caption{Nonlinear density-response spectrum $\gamma(\omega)$ for a two-site chain at $\epsilon_0=5.0$ and half-filling for selected interaction strengths $U$ in second Born approximation and using the GKBA. For the case of $U=1.0$, the crosses (and the green line) indicate the result for full, self-consistent two-time second Born calculation without applying the GKBA. In the time domain, the corresponding function $\gamma(t)$ is known to be strongly damped, compare with~\cite{puigvonfriesen09}.
 }
 \label{fig4}
\end{figure}

In Figure~\ref{fig4}, we show the nonlinear density-response $\gamma(\omega)$ in 2B+GKBA for different interaction strengths and investigate the noninteracting limit $U\rightarrow0$. For practically all values of $U$ shown, we are able to identify the double-peak structure which is absent in Hartree-Fock. This is the case, though the broadening of the peaks strongly increases with $U$. In the opposite case, for $U$ approaching zero, the double-peak structure correctly merges into a single peak indicating the degenerate transition frequencies $\omega_{\mathrm{AC}}$ and $\omega_{\mathrm{CD}}$. In the course of this, also the additional side peaks (which throughout have frequency offsets of multiples of $\Delta\omega$) vanish. In general, the resolution of the side peaks depends on the window function used in the Fourier transform of \eq{gammadefft}---while in Figure~\ref{fig3} we use a Hamming window, the data shown in Figure~\ref{fig4} are obtained by applying a Hann window. However, we note that the 
characteristic form and broadening of the peaks in 2B+GKBA is independent of the window function and also independent of $t^*$, cf.~\eq{gammadef}.

Finally, in Figure~\ref{fig4}, we compare the second Born result for $U=1.0$ also to full, self-consistent two-time NEGF calculations without the GKBA for $\beta=100$ (see the green crosses). Here, $\gamma(\omega)$ is extremely broad such that we do not obtain precise information about the dynamical properties and barely can extract frequencies for the involved transitions. The reason for this extreme broadening is a strong damping of the time-dependent occupations $\mean{\op{n}_i^\sigma}(t)$ in 2B, for details see~\cite{puigvonfriesen09}.

\section{Discussion}
\begin{figure}[t]
 \includegraphics[width=0.675\textwidth]{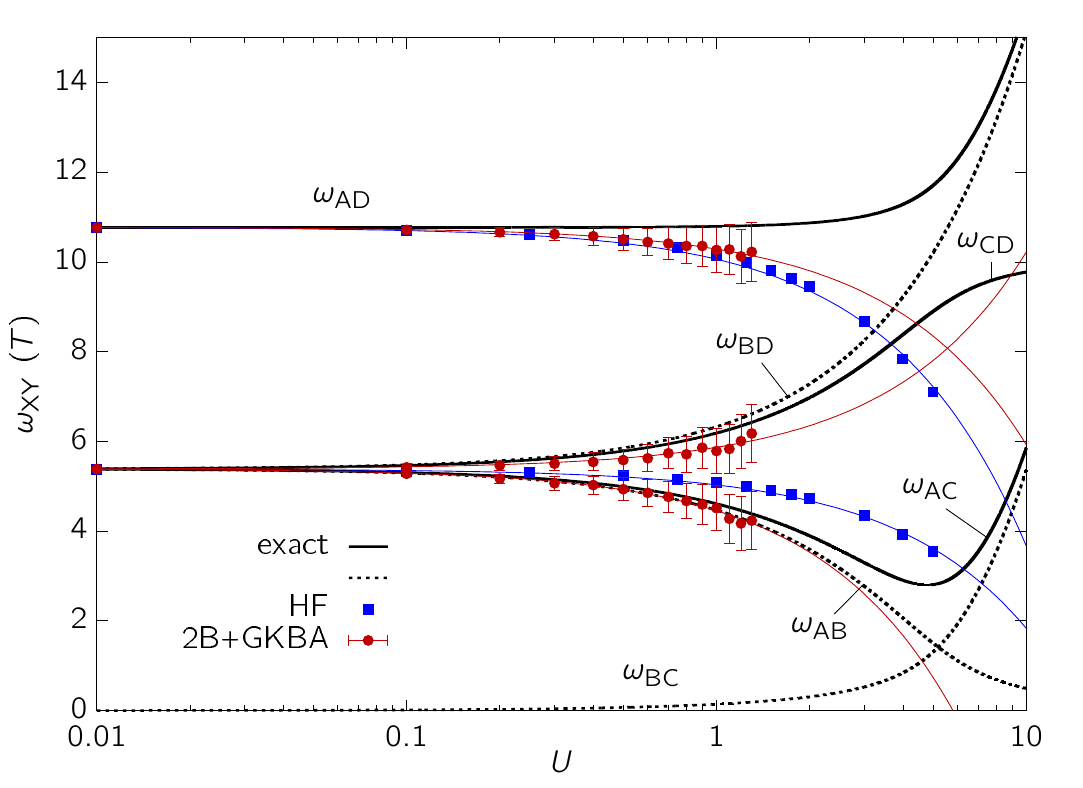}\hspace{-0.5pc}
 \includegraphics[width=0.325\textwidth]{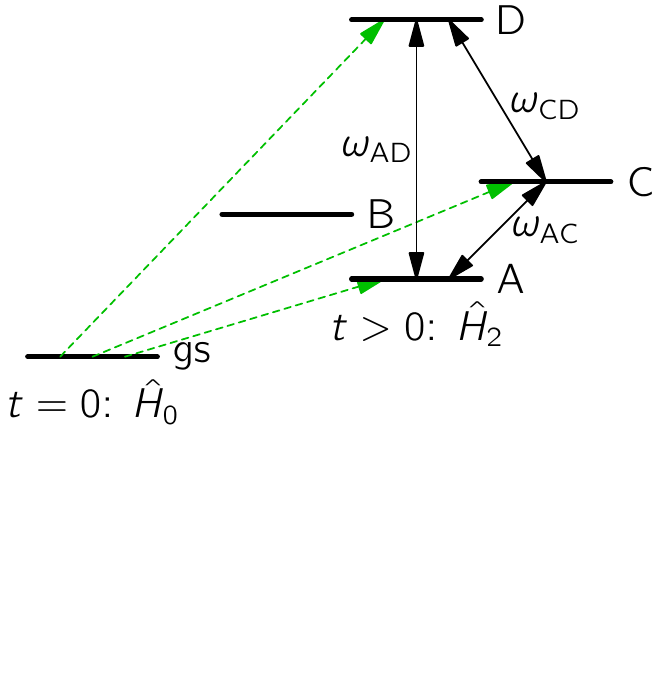}
 \caption{\underline{Left}: Comparison of the exact transition frequencies $\omega_{\mathrm{AC}}$, $\omega_{\mathrm{CD}}$ and $\omega_{\mathrm{AD}}$ (black solid lines) to the results obtained in HF (blue) and 2B+GKBA (red). The thin blue and red lines are linear fits to the data points: $\omega_{\mathrm{XY}}(U)=\alpha\,U+\omega_{\mathrm{XY},U=0}$, cf.~\eqsand{fithf}{fit2b}. \underline{Right}: Schematic view of the excitation dynamics starting from the ground state (gs) of Hamiltonian $\op{H}_0$. For $t>0$, we switch to the new Hamiltonian $\op{H}_2=\op{H}_0+\op{H}_1$, cf.~\eqsand{h0}{h1}.}
 \label{fig5}
\end{figure}

The comparison of Figure~\ref{fig1} with Figures~\ref{fig2} to \ref{fig4}, allows for the statement that 2B+GKBA seems to capture the correct physics (in contrast to the Hartree-Fock results and full 2B calculations) although there appear additional artifacts in $\gamma(\omega)$ such as broadening and side peaks. The main result is summarized in the left panel of Figure~\ref{fig5}. It shows the exact and approximate transition frequencies $\omega_{\mathrm{AC}}$, $\omega_{\mathrm{CD}}$ and $\omega_{\mathrm{AD}}$ as function of the interaction strength $U$.

In order to explain the failure of Hartree-Fock in more detail, we investigate the character of the excitation processes involved, recall the energy spectrum in the inset of Figure~\ref{fig1}~a) and see the right panel in Figure~\ref{fig5}. For the asymmetric chain with $\epsilon_0=5.0$ and $\epsilon_1=0$ described by Hamiltonian $\op{H}_2=\op{H}_0+\op{H}_1$ for $t>0$, it is easily shown~\cite{doubleexc} that the excited state $\mathrm{D}$ is a doubly-excited state relative to the ground state $\mathrm{A}$, whereas the states $\mathrm{B}$ and $\mathrm{C}$ are singly-excited states. With the same argument, state $\mathrm{D}$ is furthermore also a doubly-excited state relative to the ground state (gs) of $\op{H}_0$. This has the following consequences:
\begin{enumerate}
 \item The energetically largest transition $\mathrm{A}\leftrightarrow\mathrm{D}$ will not appear in the HF solution as double excitations are generally not included in any time-dependent Hartree-Fock calculation~\cite{balzer12_epl}.
 \item The transitions $\mathrm{A}\leftrightarrow\mathrm{C}$ and $\mathrm{C}\leftrightarrow\mathrm{D}$ are of one-electron character and therefore in principle describable by Hartree-Fock. The reason why, nonetheless, one observes only the transition $\mathrm{A}\leftrightarrow\mathrm{C}$ is that the function $\gamma(\omega)$ combined with the excitation (\ref{h1}) only probes energy differences between states that are populated already at time $t=0$. According to (i), however, the state $\mathrm{D}$ is never populated in HF.
\end{enumerate}

Points (i) and (ii), clearly explain the simplicity and shortcomings of the HF solution for the nonlinear response spectra $\gamma(\omega)$. Moreover, an analysis of the exact eigenstates of the two Hamiltonians $\op{H}_0$ and $\op{H}_2$ reveals that there is finite overlap only between the ground state of $\op{H}_0$ and the states $\mathrm{A}$, $\mathrm{C}$ and $\mathrm{D}$ of $\op{H}_2$ (all being singlets). The vanishing overlap with the eigenstate $\mathrm{B}$ (the triplet with $S_z=0$~\cite{jafari08}) is responsible for the fact that we do not observe transitions involving state $\mathrm{B}$ in the density-response. In terms of the wave function, the dynamics of the system for $t>0$ is therefore,
\begin{align}
 \ket{\Psi(t)}=c_\mathrm{A}\,\e{-\ii E_\mathrm{A}t}\ket{\mathrm{A}}+c_\mathrm{C}\,\e{-\ii E_\mathrm{C}t}\ket{\mathrm{C}}+c_D\,\e{-\ii E_\mathrm{D}t}\ket{\mathrm{D}}\;,
\end{align}
with $\ket{\mathrm{X}}$ being the eigenstates of $\op{H}_2$ (having energy $E_\mathrm{X}$) and $c_\mathrm{X}=\braket{\mathrm{X}}{\mathrm{gs}}$ denoting the expansion coefficients with respect to the ground state $\ket{\mathrm{gs}}$ of $\op{H}_0$.

In contrast to the HF approximation, a second-order self-energy (as 2B) has the ability to describe double excitations, see~\cite{sakkinen12}. For this reason, the 2B+GKBA calculations capture the transition $\omega_\mathrm{AD}$ and, accordingly, also $\omega_\mathrm{CD}$. In this context, we again emphasize that the obtained frequency for the transition $\mathrm{A}\leftrightarrow\mathrm{D}$ is not a multiple of any other frequency observed in 2B+GKBA, compare the red dots and the blue squares in the left panel of Figure~\ref{fig5}. The linear fits to the data points are,
\begin{align}
\label{fit2b}
  \omega_\mathrm{AC}^\mathrm{2B}&=-0.928\,U+5.385\;,\hspace{1pc}
    \omega_\mathrm{CD}^\mathrm{2B}=0.483\,U+5.385\;,\hspace{1pc}
      \omega_\mathrm{AD}^\mathrm{2B}=-0.385\,U+10.770\;.
\end{align}
In Hartree-Fock, we obtain,
\begin{align}
\label{fithf}
 \omega_\mathrm{AC}^\mathrm{HF}&=-0.355\,U+5.385\;,\hspace{1pc}
 \omega_\mathrm{AD}^\mathrm{HF}=-0.710\,U+10.770=2\,\omega_\mathrm{AC}^\mathrm{HF}\;.
\end{align}

With the result of \eq{fit2b}, the second Born approximation plus GKBA gives a complete picture of the nonequilibrium dynamics in the considered Hubbard chain. However, there is one additional effect in 2B+GKBA we cannot ignore. This concerns the approximate value for the double excitation frequency $\omega_{\mathrm{AD}}$ which behaves differently as function of $U$ than the values for the single excitations $\omega_\mathrm{AC}$ and $\omega_\mathrm{CD}$: While the 
approximate frequencies $\omega_{\mathrm{AC}}$ and $\omega_{\mathrm{CD}}$ follow the exact results as $U$ is increased, we observe a negative slope ($-0.385$ in the linear fit) as function of $U$ for $\omega_\mathrm{AD}$. This is opposite to the exact solution which reveals a clearly positive slope and minor absolute value of $\mathrm{d}\omega_{\mathrm{AD}}/\mathrm{d}U$ for small $U\lesssim 2$, see the upper solid black line in Figure~\ref{fig5} (left panel). 

\section{Summary}
In conclusion, we have investigated a small Hubbard-type system with respect to its excitation properties beyond the linear response regime by propagating the Kadanoff-Baym equations in HF and 2B+GKBA. In the course of this, we found clear signals of electron-electron correlations, and the absence of damped solutions for the density-response $\gamma(t)$ in 2B+GKBA allowed us to compute the approximate transition frequencies as function of the interaction strength. Comparisons with the exact Green's function furthermore showed that 2B+GKBA produces reasonable results with only minor deficiencies regarding broadening and side peak effects. Finally, our analysis underlined the importance of double excitations in finite systems.

\section*{Acknowledgments}
We thank C.~Verdozzi for a fruitful discussion and acknowledge the North-German Supercomputing Alliance (HLRN) for providing computing time.


\begin{thebibliography}{50}
\bibitem{doubleexc}{Consider for example the most simple case $U=0$, which leads to the single particle energies $E_1=-0.193$ and $E_2=5.193$. According to the inset in Figure~\ref{fig1}, the transition frequency $\omega_\mathrm{AD}$ measures $2E_2$, which means that both electrons are excited.}
\bibitem{broadening}{We note that the peak broadening is not determined by the length of the $\gamma(t)$-time series but is a property of the second Born approximation in combination with the GKBA, cf.~also Figure~\ref{fig4}.}
\end{thebibliography}
\end{document}